**Type of manuscript**: Original Articles

**Manuscript title**: Using Directed Acyclic Graphs to Illustrate Common Biases in Diagnostic Test Accuracy Studies

**Authors and affiliations**:

Yang Lu[a]; Nandini Dendukuri[a,b]

[a]Department of Epidemiology, Biostatistics and Occupational Health, McGill University, Montreal, QC, Canada

[b]Department of Medicine, McGill University, Montreal, QC, Canada

**Corresponding author**:

Nandini Dendukuri,

5252 boul. de Maisonneuve Bureau 3F.50

Montreal, Quebec, H4A 3S5

(514) 934-1934 ext: 36916

nandini.dendukuri@mcgill.ca

**Running head**: DAGs for diagnostic test accuracy studies

**Word count**: 2603

**Sources of financial support**: Natural Sciences and Engineering Research Council (NSERC) Grant number RGPIN/06713-2019

**Conflict of interest**: The authors have no conflicts of interest to disclose.

# Abstract

**Background:** Diagnostic test accuracy (DTA) studies, like etiological studies, are susceptible to various biases including reference standard error bias, partial verification bias, spectrum effect, confounding, and bias from misassumption of conditional independence. While directed acyclic graphs (DAGs) are widely used in etiological research to identify and illustrate bias structures, they have not been systematically applied to DTA studies.

**Methods:** We developed DAGs to illustrate the causal structures underlying common biases in DTA studies. For each bias, we present the corresponding DAG structure and demonstrate the parallel with equivalent biases in etiological studies. We use real-world examples to illustrate each bias mechanism.

**Results:** We demonstrate that five major biases in DTA studies can be represented using DAGs with clear structural parallels to etiological studies: reference standard error bias corresponds to exposure misclassification, misassumption of conditional independence creates spurious correlations similar to unmeasured confounding, spectrum effect parallels effect modification, confounding operates through backdoor paths in both settings, and partial verification bias mirrors selection bias. These DAG representations reveal the causal mechanisms underlying each bias and suggest appropriate correction strategies.

**Conclusions:** DAGs provide a valuable framework for understanding bias structures in DTA studies and should complement existing quality assessment tools like STARD and QUADAS-2. We recommend incorporating DAGs during study design to prospectively identify potential biases and during reporting to enhance transparency. DAG construction requires interdisciplinary collaboration and sensitivity analyses under alternative causal structures.

## Key Messages:

- Directed acyclic graphs (DAGs), widely used in etiological research to identify bias structures, have not been systematically applied to diagnostic test accuracy (DTA) studies despite their susceptibility to similar biases.
- We demonstrate that five major biases in DTA studies—reference standard error bias, misassumption of conditional independence, spectrum effect, confounding, and partial verification bias—share identical causal structures with corresponding biases in etiological studies and can be effectively illustrated using DAGs.
- Incorporating DAGs into DTA study design and reporting provides a transparent framework for both prospectively and retrospectively identifying potential biases, selecting appropriate analytical strategies, and should complement existing quality assessment tools like STARD and QUADAS-2.

## Introduction

Diagnostic test accuracy (DTA) studies are an important early step in diagnostic test evaluation. These studies estimate a diagnostic test's ability to detect the target condition of interest. Like etiologic studies, DTA studies are prone to well-recognized biases including reference standard error bias, partial verification bias, spectrum effect, confounding and the bias due to misassumption of conditional independence [1-8].

In the literature on etiologic studies, directed acyclic graphs (DAGs) are widely used to depict the causal relations between variables, to illustrate data generating processes and to investigate missing data mechanisms. In particular, DAGs are useful to identify the presence of various biases (e.g. confounding, misclassification and selection bias) that could impact estimation of causal relations [9-16]. However, to our knowledge, typical DTA studies rely on checklists like STARD and QUADAS-2 to assess and document risk of bias, and DAGs have not been widely used to depict the relevant relationships [2, 5].

In this work, our primary objective is to demonstrate how DAGs can also be useful in illustrating structures of bias commonly encountered in DTA studies. We will also show through these DAGs that the structure of bias in DTA studies and etiologic studies is similar.

## Methods

### Directed Acyclic Graphs (DAGs) for Diagnostic Test Accuracy (DTA) Studies

A DAG illustrates causal relationships using two components: vertices and directed edges. In a causal DAG, the vertices are random variables, e.g. exposure/treatment, outcome, confounder variables [9, 17]. Both observed and unobserved variables can be included in a DAG. The directed edges represent direct causal effect from one variable to another. A characteristic feature of a DAG is that it has no closed cycles.

In the context of DTA studies, we aim to evaluate how well an index test (the diagnostic test under evaluation) can identify the presence or absence of a target condition (the disease or health condition of interest). The primary variables of interest are: i) the target condition, ii) reference test result, iii) index test result, and iv) covariates or selection criteria that affect the target condition and/or the index test [18]. In practice, we may encounter more complex designs with multiple index and reference tests, covariates and selection criteria. The reference test is the best available test or method for detecting the presence of the target condition [2]. In many cases, the target condition is a latent variable that cannot be perfectly measured by any observed tests.

The primary association of interest in DTA studies is between the target condition and index test result(s). This relationship is typically quantified in terms of two separate parameters - sensitivity (P(Index test is positive | Target condition is present)) and specificity (P(Index test is negative | Target condition is absent)). When the reference standard is assumed to be perfect, the target condition and reference standard result are equivalent and can be represented as a single variable. Otherwise, we consider the target condition to be latent and treat the reference test as a distinct variable from the target condition.

## Common biases encountered in DTA studies

Table 1 summarizes the notation we will use to illustrate DAGs in DTA studies on the left and in etiological studies on the right. Variables in a given row of the table play a similar role in DTA and etiological studies, respectively.

Table 1. Variable notation in diagnostic test accuracy studies and etiological studies*

| DTA Studies | Etiological Studies |
|---|---|
| $D$ (Target Condition) | $X$ (Exposure) |
| $T_1$ (Reference test result) | $X'$ (Imperfect measurement of exposure) |
| $T_2$ (Index test result) | $Y$ (Outcome) |
| $R$ (Covariate) | $R$ (Covariate) |
| $S$ (Variable determine selection criterion) | $S$ (Variable determine selection criterion) |

\* *Variables in each row play a similar role in DTA studies and Etiological studies, respectively.*

Table 2 lists DAGs that illustrate the structure of common biases in DTA studies on the left and the corresponding bias in etiological studies (i.e. which shares the same structure) on the right. In each case, we illustrate one way in which the bias could arise using a simple design based on as few variables as possible. In the sub-sections below, we explain how each bias arises.

**Motivating examples**

For three of the biases, we use the problem of evaluating diagnostic tests for pediatric tuberculosis (PTB) as motivating example. For clarity, DAGs illustrating each bias appear in separate figures in the main text. There is no perfect reference test for pediatric tuberculosis. Diagnostic test accuracy studies typically gather information on multiple tests [19]. Traditionally, cell culture was the best available microbiological test, but its use is hampered by long turnaround times and need for advanced laboratory support [20]. The Tuberculin Skin Test, is widely used to document exposure to TB, though its sensitivity for measuring immune response is known to be affected by human immunodeficiency virus (HIV) status [21]. Newer polymerase chain reaction (PCR) tests, like GeneXpert, are also microbiological tests, which are more accessible than culture in resource-poor areas, but still have poor sensitivity [22]. Other examples are drawn from evaluating tests for Chlamydia trachomatis and Human Papilloma Virus (HPV).

Table 2. Comparison of Directed Acyclic Graphs (DAGS) illustrating the structure of common biases in etiological studies and the corresponding bias in diagnostic test accuracy Studies

| Causal DAGs for Diagnostic Test Accuracy Studies | Causal DAGs for Etiological Studies |
|---|---|
| Reference Standard Error Bias<br><br>$D \rightarrow T_2$<br>$\downarrow$<br>$T_1$ | Exposure Misclassification<br><br>$X \rightarrow Y$<br>$\downarrow$<br>$X'$ |
| Misassumption of Conditional Independence<br><br>$D \rightarrow T_2$<br>$\downarrow \quad \uparrow$<br>$T_1 \leftarrow R$ | Misclassification + Confounding<br><br>$X \rightarrow Y$<br>$\downarrow \quad \uparrow$<br>$X' \leftarrow R$ |
| Spectrum Effect | Effect Modification |

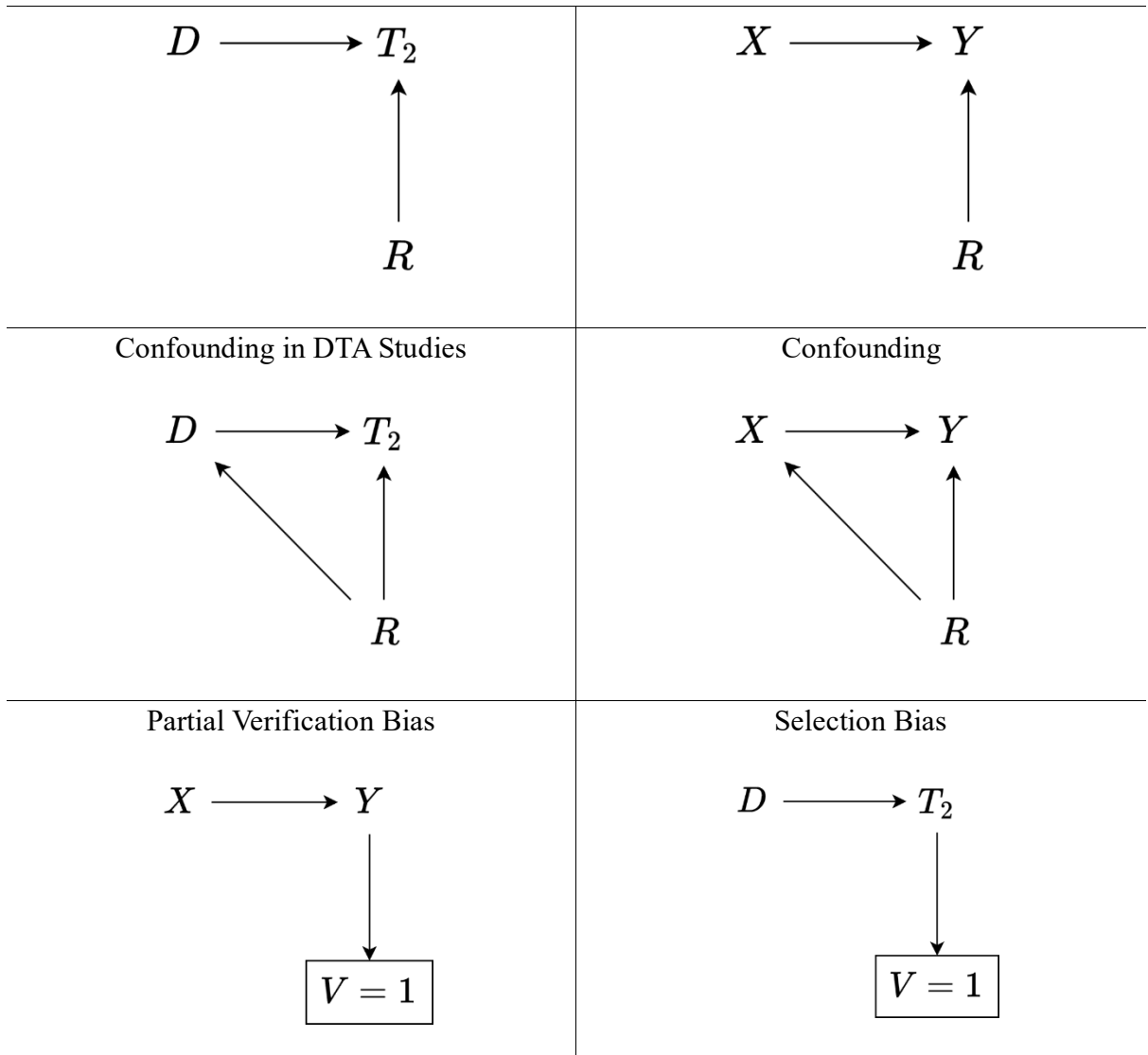

*\* See Table 1 for explanation of Notation.*

## Results

**Reference Standard Error Bias**

The reference standard error bias refers to bias in the estimated index test sensitivity and specificity when the reference test is imperfect but treated as if it were perfect when evaluating the index test accuracy [23-25].

The corresponding DAG in Table 2 illustrates the simplest possible causal structure underlying reference standard error bias. The directed edges $D \rightarrow T_1$ and $D \rightarrow T_2$ represent the true causal relationships whereby the true target condition determines the results of both diagnostic tests, depending on their respective sensitivity and specificity. Under the conditional independence assumption ($T_1 \perp T_2 \mid D$), the two test results are associated only through their shared dependence on the true target condition. When the reference test is incorrectly assumed perfect, a backdoor path arises between $T_1$ and $T_2$ via D, i.e. $T_1 \leftarrow D \rightarrow T_2$. This results in a biased estimate of the accuracy of $T_2$.

Example: Both GeneXpert (a rapid, point-of-care, polymerase chain reaction (PCR) test) and cell culture are microbiological diagnostic tests for PTB (Figure 1). When culture is incorrectly assumed to be a perfect reference test, the estimated diagnostic accuracy of GeneXpert is biased. A backdoor path is created passing through the target condition (PTB), and this backdoor path serves as a biased proxy for the true relationship between PTB and GeneXpert [26].

Equivalent bias in etiological studies: In etiological studies, this structure is recognized as information bias arising from exposure misclassification. When the true exposure X is not observable and replaced by a misclassified measure $X'$, the relation $X' \rightarrow Y$ serves as a biased approximation of the causal relation $X \rightarrow Y$ through the backdoor path ($X' \leftarrow X \rightarrow Y$) [27-29].

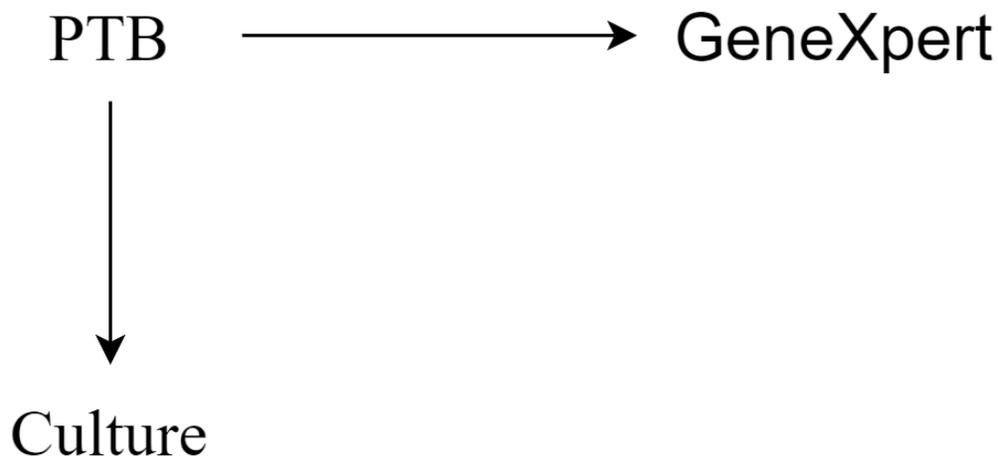

Figure 1. Backdoor path created by imperfect reference standard in pulmonary tuberculosis diagnostic test accuracy studies. PTB, pulmonary tuberculosis.

**Misassumption of Conditional Independence**

The conditional independence assumption between the reference test and index test given the subject's true target condition, i.e., $T_1 \perp T_2 \mid D,$ is commonly used in DTA studies. However, if it does not hold, it can lead to biased estimates of sensitivity and specificity. The bias can occur when we assume that the imperfect reference test is perfect (resulting in reference standard error bias) or when we adjust for the imperfect nature of the reference standard via latent class analysis (LCA) [30-32].

The DAG in Table 2 illustrates one possible mechanism that invalidates the assumption of conditional independence. In this DAG, $R$ denotes an unobserved common cause with two outgoing edges $R \rightarrow T_1$ and $R \rightarrow T_2$, demonstrating that R directly affects the two tests when conditioning on $D$. This creates a non-causal path $D \leftarrow T_1 \leftarrow R \rightarrow T_2$ that induces a spurious correlation between the target condition status and index test result beyond their causal relationship. When the conditional independence is assumed in the statistical analysis despite this structure, sensitivity and specificity estimates become biased.

Example: A real-world example is in the context of pediatric TB diagnosis, where GeneXpert ($T_2$) and culture ($T_1$) rely on tuberculosis bacterial load ($R$) in the sample (Figure 2). As a result, both tests may be falsely negative for individuals with a low bacterial load, leading them to be dependent even after conditioning on PTB status. When GeneXpert is evaluated against culture as the reference standard, failure to account for bacterial load leads to overestimation of sensitivity [26]. The bias persists within the context of a latent class analysis, a modeling approach that recognizes that culture is not perfect [26].

Equivalent in etiological studies: The misassumption of conditional independence is not commonly recognized in etiological studies to our knowledge. However, theoretically, this could arise when the exposure is misclassified and there exists a confounding factor between the misclassified exposure and the outcome.

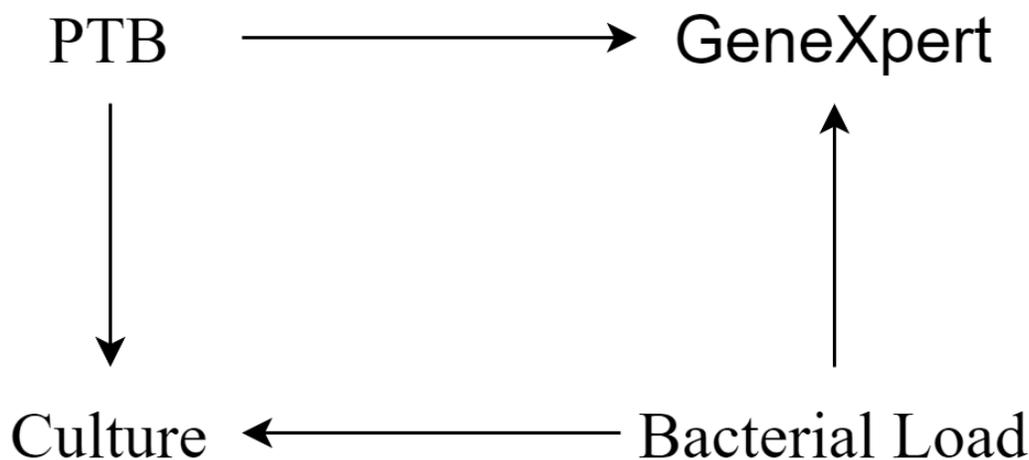

Figure 2. Conditional dependence between GeneXpert and culture through bacterial load in pulmonary tuberculosis diagnostic test accuracy studies.

**Spectrum Effect**

Spectrum effect refers to the phenomenon where the index test performs differently in different subgroups of subjects. The subgroups can be defined by patient characteristics (e.g., disease severity, age) or test characteristics, e.g. test manufacturer. It represents true variability in test performance across a spectrum and is not a bias per se [33-37].

Table 2 illustrates one possible structure that creates the spectrum effect. In this DAG, an additional variable $R$ is introduced to represent the patient or test characteristic. The edge $R \rightarrow T_2$ depicts that the patient spectrum adds extra variability to test performance, i.e., the test's sensitivity and specificity vary across levels of $R$. This extra heterogeneity in diagnostic accuracy across patient characteristics is the spectrum effect.

Example: The enzyme immunoassay for Chlamydia trachomatis diagnosis is known to have higher sensitivity and specificity in younger patients than older patients ($R$: Age $\leqslant 24$ or Age>24) [33] (Figure 3). Its performance is also likely to vary when stratified by clinic type ($R$: public family planning clinic or sexually transmitted disease clinic) - it has a higher specificity in patients suspected of sexually transmitted disease [33].

Equivalent in etiological studies: Spectrum effect can be thought of as equivalent to effect modification in a causal relationship in etiological studies, which is referred to as effect modification and is not a bias per se.

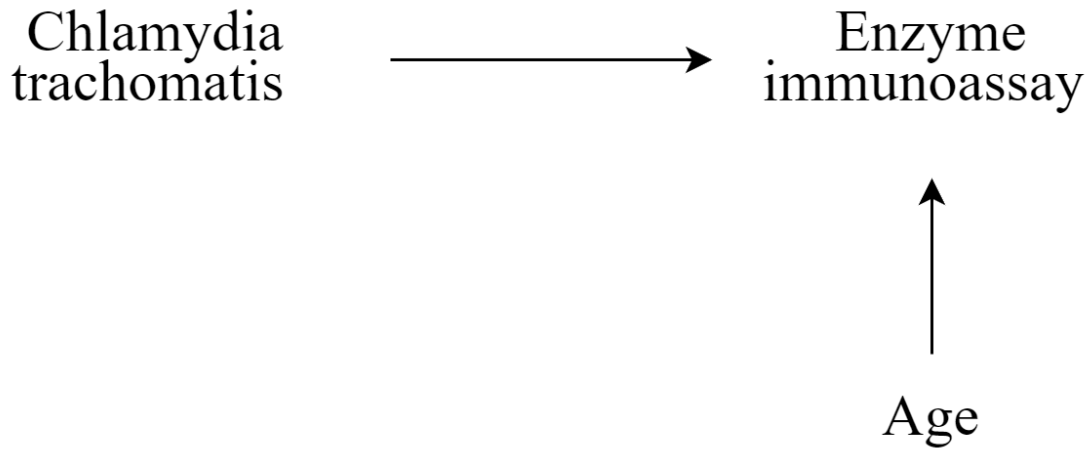

Figure 3. Spectrum effect in diagnostic test accuracy studies for Chlamydia trachomatis by age.

**Confounding**

Confounding in DTA studies occurs when a third variable influences both the target condition and the index test result, and this shared cause is not adjusted for in the statistical analysis [38, 39]. The shared cause can be any conditions related to subjects or external factors [3, 6-8].

The corresponding DAG in Table 2 illustrates one possible structure of confounding. In this DAG, we use an additional variable $R$ to represent the confounding variable with two outgoing edges $R \to D$ and $R \to T_2$, indicating this factor directly affects the target condition and the test's result. This creates a backdoor path $D \leftarrow R \to T_2$ that creates a spurious association between target condition and index test result, leading to biased estimates of sensitivity and specificity when $R$ is not adjusted for in the analysis.

Example: Positive HIV status (i.e. confounding variable $R$ = HIV) suppresses the immune response of children, reducing the accuracy of the tuberculin skin test (TST) for detecting tuberculosis infection. Meanwhile, HIV also increases the risk that a child becomes infected with tuberculosis (Figure 4). Therefore, not adjusting for HIV creates a bias in the estimated sensitivity and specificity of TST for TB infection [26].

Equivalent bias in etiological studies: The backdoor path arising due to a confounding variable is well recognized way to identify this bias in a DAG [40, 41].

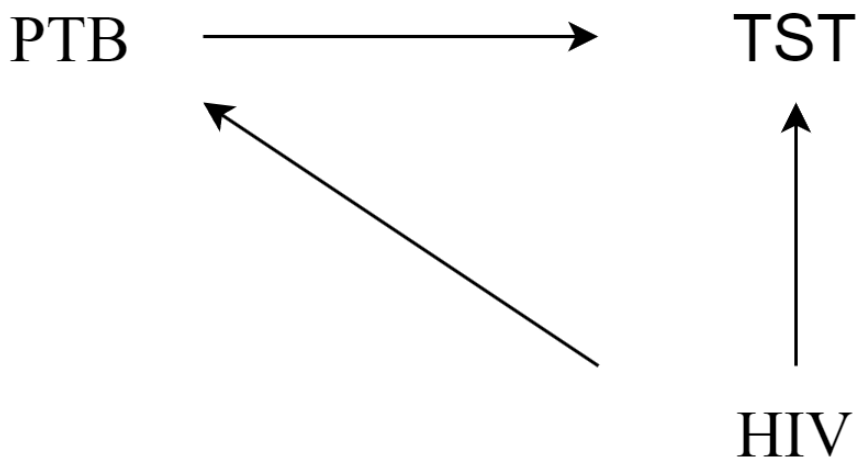

Figure 4. Confounding by human immunodeficiency virus status in pulmonary tuberculosis diagnostic test accuracy studies. PTB, pulmonary tuberculosis; TST, tuberculin skin test; HIV: human immunodeficiency viruses.

**Partial Verification Bias**

Partial verification bias arises when all patients receive the index test but only a subset also receive the reference test, and the probability of receiving the reference test (verification) depends on the result of the index test and/or on a covariate which may be related to the subject's target condition. A common example occurs when only patients with positive diagnoses by the index test (often a non-invasive or inexpensive but inaccurate test) will be tested by the reference test (which is invasive or expensive but accurate), while only a small percentage of those with negative diagnoses by the index test will be verified using the reference test [1, 35, 42]. The index test sensitivity and specificity estimated from the partially verified data are biased when the partial verification procedure is not properly adjusted for in the analysis.

The corresponding DAG in Table 2 illustrates one possible structure of partial verification bias without any covariate. In this DAG, $V$ denotes whether the reference test result is obtained. When $V$ is conditioned upon (indicated by the box around $V = 1$), the analysis is restricted to verified subjects only. This conditioning creates selection bias as the verified sample may not represent the full study population. When, selection (verification) depends on $T_2$, conditioning on $V = 1$ creates a selected sample where the $D \rightarrow T_2$ relationship (sensitivity/specificity) differs from the target population. The estimated sensitivity and specificity will be biased because they are estimated from the non-representative verified subset of subjects.

Example: A real-world example of partial verification bias involves using polymerase chain reaction (PCR) as an initial test to detect human papilloma virus (Figure 5). All subjects who tested positive on PCR received the invasive reference test (colposcopy), but only a random sample of those who were negative on PCR received it [43].

Equivalent bias in etiological studies: This structure parallels selection bias in etiological studies, where conditioning on a selection variable creates bias in exposure-outcome associations. Although typical selection bias require conditioning on a collider, it has been demonstrated that a non-random selection mechanism can alter existing association and prevents the generalizability of the effect to the target population [44, 45].

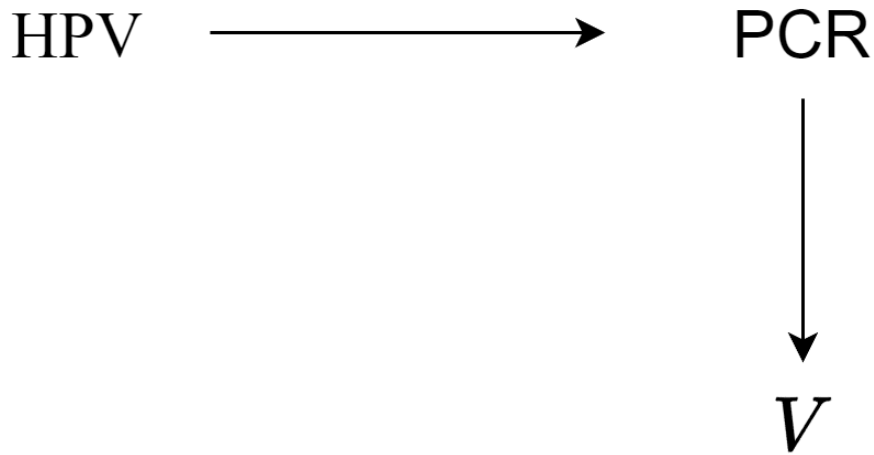

Figure 5. Partial verification bias in diagnostic test accuracy studies for human papillomavirus. HPV, human papillomavirus; PCR, polymerase chain reaction; *V*, verification variable.

## Discussion

We have demonstrated how DAGs can be helpful tools in illustrating the structure of common biases in the context of DTA studies, though our DAG for each structure represents just one of potentially many ways to conceptualize these relationships.

Through analogy with bias structures in etiological research, our work shows that most DTA biases have their well-recognized counterpart bias structures in causal inference, including exposure misclassification, confounding, effect modification and selection bias. This structural correspondence suggests that existing methodological principles developed for etiological research can inform bias assessment and suggest bias correction strategies in DTA studies. For example, the partial verification bias can be corrected using strategies stemming from selection bias correction [1, 46]. For confounding, the DAG immediately identifies the backdoor path (*D*

← $R$ → $T_2$) that must be blocked through stratification or covariate adjustment [38]. The spectrum effect can be properly studied through subgroup analysis [34].

We recommend that future DTA studies include a corresponding DAG during study design phase to prospectively identify bias sources and inform data collection strategies, or as a tool for reporting studies with known bias risk to enhance clarity. DAGs should complement existing frameworks (STARD, QUADAS-2) by providing explicit causal structures that inform interpretation of quality assessments, not replace them [2, 5]. We are cognisant that DAGs illustrate bias structure but cannot quantify magnitude nor direction. Quantitative bias analysis methods are needed to estimate whether identified biases are negligible or substantial. Inaccurate DAGs risk leading to inappropriate analytical choices. Therefore, we recommend that DAG construction should involve an inter-disciplinary team of subject-area experts and methodologist as it requires domain-specific knowledge of the target condition, test being evaluated and the causal relationships. This knowledge may be uncertain when developing novel tests or evaluating tests for new conditions. Therefore, we also recommend investigators conduct sensitivity analyses under alternative plausible structures. Though these recommended approaches are not infallible, DAGs serve to make researchers' assumptions transparent to readers for constructive feedback.

A limitation of this work is that our listing of biases is not exhaustive. The same DAGs may explain other biases we have not mentioned (e.g. selection bias due to a case-control design), and other DAGs may be needed to consider biases we have not mentioned (e.g. blinded recording of index test results). Future studies can certainly elaborate this early effort towards using DAGs for DTA studies, including DAGs for DTA studies comparing two diagnostic tests.